\documentclass[final,5p,times,twocolumn]{elsarticle}
\usepackage{amssymb}
\usepackage{amsmath}

\begin{document}

\begin{frontmatter}

\title{Program package for multicanonical simulations of U(1) lattice
gauge theory}

\author{Alexei Bazavov$^a$ and Bernd A. Berg$^b$} 

\address[UA]{Department of Physics, University of Arizona, Tucson, 
Arizona 85721}
\address[FSU]{Department of Physics, Florida State University, 
Tallahassee, Florida 32306}


\begin{abstract}
We document our Fortran~77 code for multicanonical simulations of 4D 
U(1) lattice gauge theory in the neighborhood of its phase transition. 
This includes programs and routines for canonical simulations using
biased Metropolis heatbath updating and overrelaxation, determination 
of multicanonical weights via a Wang-Landau recursion, and 
multicanonical simulations with fixed weights supplemented by
overrelaxation sweeps. Measurements are performed for the action,
Polyakov loops and some of their structure factors. Many features 
of the code transcend the particular application and are expected
to be useful for other lattice gauge theory models as well as for 
systems in statistical physics.
\medskip

\noindent {\bf Program Summary}

\noindent {\it Program title:} STMC\_U1MUCA



\noindent {\it Program obtainable from:} Temporarily from URL
{\tt http://www.hep.fsu.edu/\~\,$\!$berg/research}~.



\noindent {\it Distribution format:} tar.gz

\noindent {\it Programming language:} Fortran~77

\noindent {\it Computer:} Any capable of compiling and executing 
Fortran code.

\noindent {\it Operating system:} Any capable of compiling and executing 
Fortran code.


\noindent {\it Nature of problem:} Efficient Markov chain Monte Carlo 
simulation of U(1) lattice gauge theory close to its phase transition. 
Measurements and analysis of the action per plaquette, the specific 
heat, Polyakov loops and their structure factors.

\noindent {\it Solution method:} Multicanonical simulations with an 
initial Wang-Landau recursion to determine suitable weight factors.
Reweighting to physical values using logarithmic coding and calculating
jackknife error bars.

\noindent {\it Running time:} The prepared tests runs took up to 
74 minutes to execute on a 2GHz~PC.

\end{abstract}



\end{frontmatter}


\section{Introduction}

Continuum, quantum gauge theories are defined by their Lagrangian 
densities, which are functions of fields living in 4D Minkowski 
space-time. In the path-integral representation physical observables 
are averages over possible field configurations weighted with an 
exponential factor depending on the action. By performing a Wick 
rotation to imaginary 
time the Minkowski metric becomes Euclidean. Discretization of this 
4D Euclidean space results in lattice gauge theory (LGT) -- a 
regularization of the original continuum theory, which allows to 
address non-perturbative problems. For a textbook see, for instance,
Ref.~\cite{Rothe}. Physical results are recovered in the quantum 
continuum limit $a\to 0$, where $a$ is the lattice spacing measured
in units proportional to a diverging correlation length.

U(1) pure gauge theory, originally introduced by Wilson~\cite{Wi74},
is a simple 4D LGT. Nevertheless, determining its phase structure 
beyond reasonable doubt has turned out to be a non-trivial 
computational task. One encounters a phase transition which is 
believed to be first-order on symmetric $N_s^4$ lattices, e.g. 
\cite{JeNe83,ArLi01,ArBu03}. For a finite temperature $N_s^3\times 
N_{\tau}$, $N_{\tau}<N_s$ geometry the situation is less clear: Either 
second-order for small $N_{\tau}$ and first-order for large $N_{\tau}$ 
\cite{VeFo04}, or always second order, possibly corresponding to a 
novel renormalization group fixed point~\cite{BeBa06}. In 3D U(1) 
gauge theory is confining for all values of the coupling on symmetric 
$N_s^3$ lattices \cite{Polyakov:1976fu,Gopfert:1981er}, while in
the finite temperature $N_s^2\times N_{\tau}$, $N_{\tau}<N_s$ geometry 
a deconfining transition of Berezinsky-Kosterlitz-Thouless type
\cite{Berezinsky:1970fr,Kosterlitz:1973xp} is expected, see 
\cite{Borisenko:2008sc} for recent numerical studies.

In lattice gauge theory one can evaluate Euclidean path integrals 
stochastically by generating an ensemble of field configurations 
with Markov chain Monte Carlo (MCMC). In this paper we report MCMC 
techniques we used in~\cite{BeBa06}. They are based on multicanonical 
(MUCA) simulations~\cite{BeNe91,Bbook} supplemented by a Wang-Landau 
(WL) recursion \cite{WaLa01}, both employed in continuum formulations. 
For updating we use the biased Metropolis heatbath algorithm (BMHA) 
of~\cite{BaBe05} added by overrelaxation~\cite{Ad81}. Observables 
include the specific heat, Polyakov loops and their structure factors 
(SFs) for low-lying momenta. For the analysis of these data binning
is used to render autocorrelations negligible and a logarithmically 
coded reweighting procedure calculates averages with jackknife error 
bars. 

Our program package 
$$ {\tt STMC\_U1MUCA.tgz} $$
can be downloaded from the web at \smallskip

\centerline{\tt http://www.hep.fsu.edu/\~\,\!berg/research\ .}
\smallskip

Unfolding of {\tt STMC\_U1MUCA.tgz} with\ \ \ {\tt tar\ -zxvf}\ \ 
\ creates a tree structure with root directory {\tt STMC\_U1MUCA}. 
Folders on the first
level are {\tt ExampleRuns}, {\tt ForProg} and {\tt Libs}. Besides in
the subfolders of {\tt ExampleRuns}, copies of all main programs are
found in {\tt ForProg}.  Fortran functions and subroutines of our code 
are located in the subfolders of {\tt Libs}, which are {\tt Fortran}, 
{\tt Fortran\_par}, {\tt U1}, and {\tt U1\_par}.  Routines in {\tt 
Fortran} and {\tt U1} are modular, so that they can be called in any 
Fortran program, while routines in the other two subfolders need user 
defined parameter files, which they include at compilation. General 
purpose routines are in the {\tt Fortran} subfolders and to a large
extent taken over from {\tt ForLib} of \cite{Bbook}, while routines 
specialized to U(1) are in the {\tt U1} folders. Parameter files are 
\begin{equation} \label{par_cano}
  \tt bmha.par,\ lat.par,\ lat.dat,\ mc.par 
\end{equation} 
for canonical simulations and in addition
\begin{equation} \label{par_muca}
  \tt sf.par,\ u1muca.par
\end{equation} 
for SF measurements and MUCA simulations with WL recursion. The 
main programs and the routines of the subfolders {\tt Fortran\_par} 
and {\tt U1\_par} include common blocks when needed. These common 
blocks with names {\tt common*.f} are also located in the {\tt 
Fortran\_par} and {\tt U1\_par} subfolders.

This paper is organized as follows. In Sec.~\ref{sec_U1cano} we define
U(1) LGT and introduce the routines for our BMHA and for measurements 
of some observables. Sec.~\ref{sec_U1MUCA} is devoted to our code for 
MUCA runs and to the analysis of these data. Sections~\ref{sec_U1cano} 
and~\ref{sec_U1MUCA} both finish with explicit example runs, where 
Sec.~\ref{sec_U1MUCA} uses on input action parameter estimates obtained 
in Sec.~\ref{sec_U1cano}. A brief summary and conclusions are given in 
the final section~\ref{sec_conclusions}.

\section{Canonical simulations \label{sec_U1cano}}

Our code is written for a variable dimension $d$ and supposed to
work for $d\ge 2$. However, its use has mainly been confined to
$d=4$ to which we restrict most of the subsequent discussion.

After U(1) gauge theory is discretized its fundamental degrees of 
freedom reside on the links of a 4D hypercubic lattice which we label 
by $x,\mu$: $x$ is a 4D vector giving the location of a site, and $\mu
=1,2,3,4$ is the direction of the link originating from this site, 
$\mu=4$ corresponds to the temporal direction of extension $N_\tau$ 
and $\mu=1,2,3$ to the spatial directions of extension $N_s$.

The system contains $N_s^3\times N_\tau$ sites and $N_s^3\times 
N_\tau\times 4$ degrees of freedom $U_{x,\mu}$ that belong to U(1) 
gauge group, which we parametrize by
\begin{equation}\label{Utheta}
  U_{x,\mu}=\exp(i\theta_{x,\mu}),\,\,\,\,\,
  \theta_{x,\mu}\in[0,2\pi)\ .
\end{equation}
In our code we use Wilson's action
\begin{equation}\label{WiPlAct}
  S=\sum_x\sum_{\mu=1}^{4}\sum_{\nu<\mu} {\rm Re}
  (U_{x,\mu}U_{x+\hat\mu,\nu}U^+_{x+\hat\nu,\mu}U^+_{x,\nu})\ .
\end{equation}
The product $U_{x,\mu}U_{x+\hat\mu,\nu}U^+_{x+\hat\nu,\mu}U^+_{x,\nu}$ 
is taken around a \textit{plaquette}, an elementary square of the 
lattice. In 4D each link participates in 6 plaquettes ($2\,(d-1)$ in 
$d$-dimension). Products such as $U_{x+\hat\mu,\nu}U^+_{x+\hat\nu,\mu}
U^+_{x,\nu}$ are called \textit{staples}.

In canonical simulations one generates an ensemble of configurations 
weighted with $\exp(\beta S)$, the Boltzmann factor of a system with 
energy $-S$, which is in contact with a heatbath at inverse temperature 
$\beta$. Here $\beta$ is the inverse temperature of a statistical 
mechanics on the lattice and not the physical temperature of the 
LGT. The latter is given by the temporal extent of the lattice: 
$T=1/(aN_\tau)$.

An important property of the action (\ref{WiPlAct}) is \textit{locality}: 
For a link update only its interaction with a small set of neighbors is 
needed. The part of the action involving a link $U_{x,\mu}$ being updated
(with all other links frozen) is:
\begin{eqnarray}\label{ActLink}
  S(U_{x,\mu})&=& {\rm Re}\left\{U_{x,\mu}\left[
  \sum_{\nu\neq\mu}U_{x+\hat\mu,\nu}U^+_{x+\hat\nu,\mu}U^+_{x,\nu}
  \right.\right.\nonumber\\
  &+&\left.\left.\sum_{\nu\neq\mu}
  U^+_{x+\hat\mu-\hat\nu,\nu}
  U^+_{x-\hat\nu,\mu}U_{x-\hat\nu,\nu})\right]\right\}.
\end{eqnarray}
The sum in square brackets $[...]$ runs over 6 staples and is evaluated
before updating the link. We denote it
\begin{eqnarray}\label{SumSt}
  & U_\sqcup & =~\ \alpha\exp(i\theta_\sqcup)\\ \nonumber &=&\left[
  \sum_{\nu\neq\mu}U_{x+\hat\mu,\nu}U^+_{x+\hat\nu,\mu}U^+_{x,\nu} +
U^+_{x+\hat\mu-\hat\nu,\nu}U^+_{x-\hat\nu,\mu}U_{x-\hat\nu,\nu}\right].
\end{eqnarray}
To simplify the notation we drop the $x,\mu$ subscripts of the link:
\begin{equation}
  S(U)\sim {\rm Re}(UU_\sqcup).
\end{equation}
Thus the distribution
\begin{equation}
  P(U)\sim e^{\,\beta S}=e^{\,\beta{\rm Re}(UU_\sqcup)}
\end{equation}
needs to be sampled. In angular variables
\begin{equation}
  {\rm Re}(UU_\sqcup)=\alpha{\rm Re}\left(
  e^{i(\theta+\theta_\sqcup)}\right)=\alpha\cos(\theta+\theta_\sqcup)
\end{equation}
and
\begin{eqnarray}
  P(\theta)d\theta\sim e^{\,\beta\alpha\cos(\theta+\theta_\sqcup)}
  d\theta=e^{\,\beta\alpha\cos(\varphi)}d\varphi,~~
  \varphi=\theta+\theta_\sqcup.\nonumber
\end{eqnarray}
The final probability density function (PDF) is
\begin{equation}\label{Pphi}
  P(\alpha;\varphi)\sim e^{\,\beta\alpha\cos\varphi}.
\end{equation}
It is straightforward to implement the Metropolis algorithm \cite{Me53} 
for (\ref{Pphi}). A value $\varphi_{new}$ is proposed uniformly in the 
interval $[0,2\pi)$ and then accepted with probability
\begin{equation}\label{Pmet}
  \min\left\{1,\frac{P(\alpha;\varphi_{new})}
  {P(\alpha;\varphi_{old})}\right\}\,.
\end{equation}
However, it has low acceptance rate
in the region of interest ($0.8\leqslant\beta\leqslant1.2$). An 
efficient heatbath algorithm (HBA) is hard to design since the 
cumulative distribution function (CDF)
\begin{equation}\label{FPphi}
  F_P(\alpha;\varphi)=
  \frac{\int_0^\varphi P(\alpha;\varphi')d\varphi'}
  {\int_0^{2\pi} P(\alpha;\varphi')d\varphi'}
\end{equation}
is not easily invertible, because it cannot be represented in terms 
of elementary functions. Nevertheless, two variations of heatbath 
algorithms for U(1) do exist \cite{We89}, \cite{HaNa92}.

\begin{figure}
\includegraphics[width=\columnwidth]{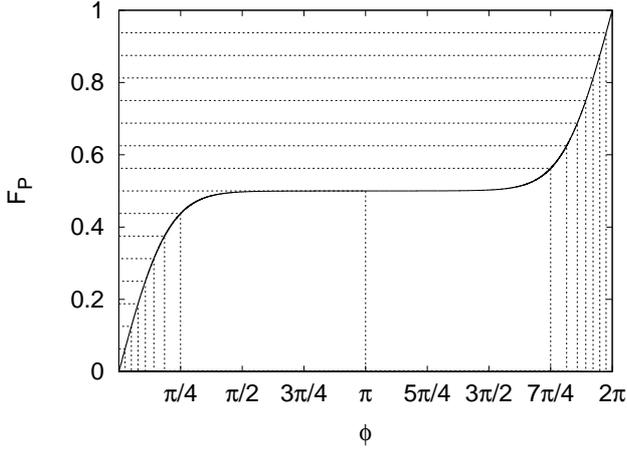}
\caption{The cumulative distribution function $F_P(\alpha=3.9375;\varphi)$
for an arbitrarily chosen value of the parameter $\alpha$.}
\label{fig_FP}
\end{figure}

\subsection{Biased Metropolis heatbath algorithm}

The MCMC updating of our code relies on a BMHA \cite{BaBe05}, which 
approximates heatbath probabilities by tables as described in the 
following. The updating variable $\varphi$ is drawn from some 
distribution $Q(\alpha;\varphi)$ and then accepted with probability
\begin{equation}\label{PQmet}
  \min\left\{1,\frac{P(\alpha;\varphi_{new})}
  {P(\alpha;\varphi_{old})}\,
  \frac{Q(\alpha;\varphi_{old})}
  {Q(\alpha;\varphi_{new})}\right\}\,.
\end{equation}
This turns out to be a special case of general acceptance probabilities 
introduced by Hastings \cite{Ha70}. One refers to to the proposals as 
\textit{biased} when $Q(\alpha;\varphi_{old})/Q(\alpha;\varphi_{new})
\ne 1$ holds. 

\begin{figure}
\includegraphics[width=\columnwidth]{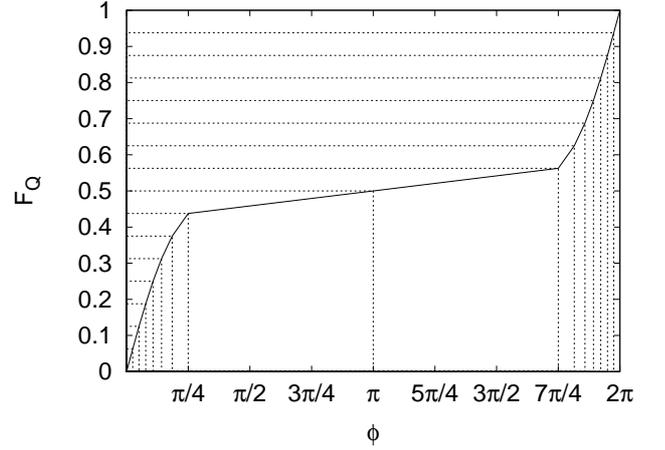}
\caption{The cumulative distribution function $F_Q(\alpha=3.9375;\varphi)$
which serves as an approximation of $F_P$. Compare to Fig.~\ref{fig_FP}.}
\label{fig_FQ}
\end{figure}

For the heatbath algorithm the proposal probability $Q(\alpha;\varphi)$
is chosen to be identical to the target distribution $P(\alpha;\varphi)$,
so that $\varphi_{new}$ is always accepted. In practice it is sufficient 
that $Q(\alpha;\varphi)$ is a good approximation of $P(\alpha;\varphi)$.
We approximate the CDF (\ref{FPphi}) by a piece-wise linear function 
$F_Q(\alpha;\varphi)$. Compare Fig.~\ref{fig_FP} and Fig.~\ref{fig_FQ}. 
We partition the $F_Q$ axis into $n$ equidistant pieces ($n=16$ in 
the figures), which defines $\varphi^1,...,\varphi^n$ values via the 
relation $F_Q(\alpha;\varphi^j)-F_Q(\alpha;\varphi^{j-1})=1/n$, and
we call an interval $(\varphi^{j-1},\varphi^j]$ a bin. The approximated
PDF 
\begin{equation}\label{Qmet}
  Q(\alpha;\varphi) = \frac{dF_Q}{d\varphi} =
  \frac{1}{n(\varphi^j-\varphi^{j-1})}
  = \frac{1}{n\Delta\varphi^j} 
\end{equation}
is flat in each bin and it is easy to generate $\varphi$ from 
$Q(\alpha;\varphi)$: One first picks a bin $j$ with uniform probability 
($1/n$) and then generates $\varphi$ uniformly in this bin.

We call this scheme BMHA because a Metropolis accept/reject step is used 
and a bias is introduced to enhance the acceptance rate by making the 
proposal probability an approximation of the one of the heatbath 
algorithm. When the discretization step goes to zero 
\begin{equation}
  \frac{\Delta\varphi^{j_{new}}}{\Delta\varphi^{j_{old}}}
  \to\frac{1/P(\alpha;\varphi_{new})}{1/P(\alpha;\varphi_{old})}=
  \frac{P(\alpha;\varphi_{old})}{P(\alpha;\varphi_{new})}
\end{equation}
follows from (\ref{Qmet}) and the acceptance rate (\ref{PQmet}) 
approaches~1. 

$P(\alpha;\varphi)$ depends on an external parameter $0\leqslant\alpha
\leqslant \alpha_{max}=2\,(d-1)$ and the inverse temperature $\beta$. 
We discretize $\alpha$ for the proposal probability $Q(\alpha;\varphi)$. 
Before updating a link in the code we evaluate the sum of the staples 
$U_\sqcup$ and decompose it into the magnitude $\alpha$ and phase 
$\theta_\sqcup$. At this stage $\alpha$ is then known.

The following is a summary of the algorithm we use to generate the
$\varphi$ variable with the PDF (\ref{Pphi}). These routines are 
located in {\tt Libs/U1\_par}.

\subsubsection{Table generation}

\begin{enumerate}

  \item Choose $m$ bins for the parameter axis $\alpha$ and $n$ bins 
  for the variable $\varphi$. Two $m\times n$ arrays are needed. We 
  take $m$ and $n$ to be powers of 2, because $n$ being a power of 
  2 speeds up finding $j_{old}$ (\ref{jold}), though $m$ can, in 
  principle, be arbitrary.

  \item Calculate a discrete set of $\alpha$ values by
  \begin{equation}
    \alpha^i=\left(i-\frac{1}{2}\right)\,
    \frac{\alpha_{max}}{m},\,\,\,\,\,i=1,...,m.
  \end{equation}

  \item For each $\alpha^i$ evaluate
  \begin{equation}
      F_Q^i(\varphi)=
      \frac{\int_0^\varphi P(\alpha^i;\varphi')d\varphi'}
      {\int_0^{2\pi} P(\alpha^i;\varphi')d\varphi'}
  \end{equation}
  by numerical integration with $\beta = \texttt{beta\_table}$ as 
  set in \texttt{mc.par}. The inverse temperature of the canonical 
  simulation is denoted \texttt{beta} in the code. We reserve the 
  possibility to have different values of the inverse temperature
  for the BMHA table generation and for the simulation. Of course, 
  for $\tt beta\_table=beta$ the table approximates the heatbath 
  distribution at {\tt beta}.
  When a range of inverse temperatures is used, as in MUCA simulations, 
  one can tune the acceptance rate by playing with \texttt{beta\_table}.
  The ranges that we used in multicanonical runs were narrow, so we 
  were content with setting $\tt beta\_table = bmax$.

  \item Tabulate $F_Q^i(\varphi)$ by values $\varphi^{i,j}$ defined by
  \begin{equation}
    \frac{j}{n}=F_Q^i(\varphi^{i,j})~~\Leftrightarrow~~
    \varphi^{i,j}=\left(F_Q^i\right)^{-1}
    \left(\frac{j}{n}\right)
  \end{equation}
  and the differences
  \begin{equation}
    \Delta\varphi^{i,j}=\varphi^{i,j}-\varphi^{i,j-1}\ .
  \end{equation}
\end{enumerate}
The common block {\tt common\_bmhatab.f}, which is listed below,
stores the quantities $\varphi^{i,j}$, $\Delta\varphi^{i,j}$
and $\ln\Delta\varphi^{i,j}$, respectively:
\begin{verbatim}
C Table for BMHA.
      common/tabbma/ tabbm(nbm2,nbm1),
     &   delbm(nbm2,nbm1),dbmln(nbm2,nbm1)
\end{verbatim}
where the parameters ${\tt nbm1}=m$ and ${\tt nbm2}=n$ are set in 
the file \texttt{bmha.par}:
\begin{verbatim}
c Biased Metropolis-HeatBath (BMHA) parameters:
      parameter(nbm1=2**5,
     &   n2log2=7,nbm2=2**n2log2) 
\end{verbatim}

\subsubsection{BMH updating}

Our implementation of BMH updates is given in the routine {\tt 
u1\_bmha.f}. A call to {\tt u1\_bmha.f} performs one \textit{sweep}, 
which here is defined by updating each variable (U(1) matrix) once 
in sequential order. For each, single BMH update it calls the 
subroutine {\tt u1\_bmha\_update.f}. Its functions are shortly 
described in the following. 

\begin{enumerate}

  \item After $\alpha$ and $\theta_\sqcup$ are calculated by the
  \texttt{u1\_getstaple} subroutine, determine the $\alpha$ bin 
  $k={\rm Int} [\alpha/\alpha_{max}]+1$, where ${\rm Int}[x]$ 
  denotes rounding to the largest integer $\leqslant x$.

  \item For given $k$ determine to which bin the previous value 
  $\varphi_{old}=(\theta_{old}+\theta_\sqcup)\mod 2\pi$ belongs 
  (in the code $\theta_{old}$ value is stored in \texttt{aphase} 
  array). This is done with a binary search
  \begin{equation} \label{jold}
    j_{old}\to j_{old}+2^i\cdot sign(\varphi-\varphi^{i,j_{old}}),
    ~~i\to i-1
  \end{equation}
  starting with $j_{old}=0$, $i=\log_2n-1$.

  \item Pick a new bin 
  \begin{equation}
    j_{new}={\rm Int}[nr_1]+1\,,
  \end{equation}
  where $r_1$ is a uniform random number in [0,1).

  \item Pick a new value
  \begin{equation}
    \varphi_{new}=\varphi^{i,j_{new}}-
    \Delta\varphi^{i,j_{new}}r_2\,,
  \end{equation}
  where $r_2$ is a uniform random number in [0,1).

  \item Accept $\varphi_{new}$ with probability 
        (\ref{PQmet}).
  \item If accepted, store 
  $\theta_{new}=(2\pi+\varphi_{new}-\theta_\sqcup)\mod2\pi$ in 
  the \texttt{aphase} array.

\end{enumerate}

For U(1) and SU(2) we found \cite{BaBe05} that $m=32$ and $n=128$ are 
large enough to achieve $\sim 0.97$ acceptance rate. Thus, the BMHA 
achieves practically heatbath efficiency. It becomes important for 
cases where a conventional HBA is difficult to implement and/or 
computationally inefficient.

\subsection{Overrelaxation}

We use overrelaxation (OR) to faster decorrelate the system. OR
algorithms were introduced by Adler. See \cite{Ad88} and references
given therein. In the formulation of Ref.~\cite{Cr87} the 
idea is to generate a new value of the link matrix that lies as far 
as possible away from the old value without changing the action too 
much. This is done by reflecting the old matrix about the link matrix, 
which  maximizes the action locally. In U(1) LGT one reflects 
$\theta_{old}$ about the element $\theta_0$, which maximizes 
the PDF~(\ref{Pphi}). The $\varphi$ value that maximizes 
(\ref{Pphi}) is $\varphi_0=\theta_0+\theta_\sqcup$. Reflecting 
$\theta_{old}$ about $\theta_0=-\theta_\sqcup$ we find
\begin{equation}\label{thetaOR}
  \theta_{new}=\theta_0-(\theta_{old}-\theta_0) =
  -2\theta_\sqcup-\theta_{old}\,.
\end{equation}
As $\theta_{new}$ (\ref{thetaOR}) does not change the action, OR 
constitutes in our case a \textit{microcanonical} update. Our 
implementation is the subroutine {\tt u1\_over.f}, which performs 
one overrelaxation sweep.  In the code $\theta_{new}=
(6\pi-2\theta_\sqcup-\theta_{old})\mod 2\pi$.

\subsection{Example runs}

Short canonical simulations are needed to determine the action range 
for the multicanonical runs. We perform 1~BMHA sweep followed by 
2~OR sweeps. Runs in 3D on $6^2\times 4$ lattices are prepared in 
the subfolders
$${\tt C3D04t06xb1p0}~~{\rm and}~~{\tt C3D04t06xb2p0}~\ $$
of the folder {\tt ExampleRuns} and in 4D on $6^3\times 4$ lattices in
$${\tt C4D04t06xb0p9}~~{\rm and}~~{\tt C4D04t06xb1p1}\ . $$
Parameters are set in the {\tt *.par} and {\tt lat.dat} files, the
lattice size in  {\tt lat.par} and {\tt lat.dat}: Run parameters
in {\tt mc.par} and the BMHA table size in {\tt bmha.par}, which
is kept the same for all runs. 

The general structure of our MCMC simulations is that outlined in 
\cite{Bbook}: Lattice set-up and initialization are done by the routine 
{\tt u1\_init.f} followed by {\tt nequi} sweeps for equilibration, 
which do not record measurements. Afterwards $\tt nrpt\times nmeas$ 
measurements are carried out, each after {\tt nsw} sweeps (in 
\cite{Bbook} only $\tt nsw=1$). 

The 4D $\beta$ values embrace the pseudo-transition region of the
$6^3\times 4$ lattice: $\beta=0.9$ in the disordered and $\beta=1.1$ 
in the ordered phase. To avoid divergence of the equilibration time
with increasing lattice size, the start configuration has to match
the phase: Ordered ($\tt istart=1$) in the ordered and disordered
($\tt istart=2$) in the disordered phase.

The production program has to be compiled with
$$\tt ../make77\ u1\_ tsbmho.f$$
where the {\tt make77} file is located one level up in the tree. Besides 
Fortran~77 compilation the {\tt make77} moves parameter files around, 
so that they are properly included by subroutines, which need them. The 
Fortran compiler defined in our {\tt make77} is g77. You may have to 
replace it with the one used on your computing platform. In the tcshell 
CPU times are recorded by running the {\tt a.out} executable with
$$\tt ../CPUtime\ >\&\ fln.txt\ \& $$
in the background, where the file {\tt CPUtime} is also located one 
level up in the tree. While running one can monitor the progress by 
looking up the {\tt progress.d} file. 
In test runs on a Intel E5405 2~GHz quad-core PC using g77 version 
3.4.6 (Red Hat 3.4.6-4) the execution times for the prepared 
simulations were 3m35s in 4D and 41s in 3D.

After a production run {\tt ana\_ts1u1.f} is used to analyze the action 
data. In the present context we only need the mean action values as 
input for the action ranges set in our multicanonical runs. With {\tt 
gnuplot h01.plt} or {\tt i01.plt} one plots the action histogram. The 
BMHA acceptance rate is also returned by {\tt ana\_ts1u1.f}.  Further, 
the integrated autocorrelation length $\tau_{\rm int}$ is estimated, 
but the statistics may not always be sufficiently large for a good 
determination (the statistics needed to estimate the average action 
is much smaller \cite{Bbook}). The relevant parameters for calculating 
$\tau_{\rm int}$ have to be set in the analysis program and a plot is 
obtained with gnuplot {\tt a1tau.plt}.

As rounding errors in floating point operations depend on the compiler,
the actual numbers obtained for the average action on different computing
platforms may vary within statistical errors. In our 4D run they were
\begin{eqnarray} 
  {\tt actm/np} &=& 0.46942\,(18)~~{\rm at}~~\beta=0.9\,,\label{actm0p9}\\
  {\tt actm/np} &=& 0.71728\,(14)~~{\rm at}~~\beta=1.1\,,\label{actm1p1}
\end{eqnarray}
where $n_p$ is the number of plaquettes. Approximated, this range across 
the phase transition will be used in the {\tt u1muca.par} file of our 
subsequent MUCA simulation:
\begin{eqnarray} \label{actmin}
  {\tt actmin}=S_{\min} &=& {\tt actm}~~{\rm at}~~\beta_{\min}=0.9\,, \\
  {\tt actmax}=S_{\max} &=& {\tt actm}~~{\rm at}~~\beta_{\max}=1.1\,. 
\label{actmax} \end{eqnarray}

The procedure for running 3D examples is the same as for the 4D runs 
described above. To cover the pseudo-transition region on $6^2\times 
4$ lattice we set the inverse temperatures for the canonical runs at 
$\beta=1.0$ in the disordered and $\beta=2.0$ in the ordered phase. 
The choice of a broader range of temperatures is appropriate, related 
to finite size effects that scale logarithmically with the lattice 
size. Average action values obtained are
\begin{eqnarray}
  {\tt actm/np}&=&0.47495\,(27)~~{\rm at}~~\beta=1.0\,,\label{actm1p0}\\
  {\tt actm/np}&=&0.81079\,(15)~~{\rm at}~~\beta=2.0\,.\label{actm2p0}
\end{eqnarray}

\section{Multicanonical simulations} \label{sec_U1MUCA}

In MUCA \cite{BeNe91} simulations one samples the phase space with 
weights, which are a {\it working estimate} \cite{Bbook} of the 
inverse spectral density up to an overall factor. This makes the 
energy histogram approximately flat and allows for efficient 
reconstruction of canonical expectation values in a range of
temperatures $(\beta_{min},\beta_{max})$. This has many applications
and is especially useful when studying the vicinity of first-order 
phase transitions, where canonical histograms exhibit a double-peak 
structure, suppressing tunneling between phases even for relatively 
small-sized system (e.g., see \cite{BBD08} for a recent study of 3D 
Potts models). For second-order phase transitions the usefulness of 
MUCA simulations has been discussed in the context of cluster 
algorithms~\cite{BeJa07}. Most important and successful applications 
target complex systems like, for instance, biomolecules~\cite{Ya08}.

The MUCA method consists of two parts \cite{Bbook}: A recursion to 
determine the weights and a simulation with frozen weights. For the 
first part we have programmed a continuous version of the WL recursion 
\cite{WaLa01}.

\subsection{Wang-Landau recursion}

Because the WL code of this section is programmed for generic use in
statistical physics systems we use the energy $E$ instead of the
action notation. In earlier recursions for MUCA parameters (see, e.g., 
\cite{Bbook}) one was iterating with a weight function of the energy, 
$w(E)$, inversely proportional to the number of histogram entries at 
$E$. In contrast to that the WL algorithm \cite{WaLa01} increments 
the weight multiplicatively, i.e., additively in logarithmic coding:
\begin{equation} \label{WLupd}
  \ln w(E{''}) \to \ln w(E^{''}) - a_{WL}\,,~~a_{WL}>0
\end{equation}
at every WL update attempt $E\to E'$, where ${\tt addwl}= a_{WL}$ in 
our code. Here $E^{''}=E$ when the update is rejected and $E^{''}=E'$ 
when the update is accepted. After a sufficiently flat histogram is 
sampled (we come back to this point), the WL parameter is refined,
\begin{equation} \label{WLrec}
  a_{WL} \to a_{WL}/2\ .
\end{equation}
In its original version the WL algorithm deals with discrete systems 
like Ising or Potts models for which histogram entries correspond 
naturally to the discrete values  of the energy. However, for 
continuous systems binsizes are free parameters and we have to deal
with their tuning.

We discretize the U(1) action range into bins of a  size large enough
in $\triangle S$, so that one update cannot jump over a bin. Here
$\triangle S < 4\,(d-1)$, because there are $2\,(d-1)$ plaquettes 
attached to a link and the range of the plaquette action is 
in the interval $[-1,+1]$, which gives another factor~2. In the program 
this value $\triangle S$ is defined as {\tt delE}. Next, we are {\it not} 
using constant weights over each bin, but instead a constant 
temperature interpolation as already suggested in \cite{BeNe91}. 

One WL update has two parts:
\begin{enumerate}
  \item A MUCA update of the energy (in our U(1) code action) using 
        the weights at hand.
  \item A WL update (\ref{WLupd}) of the MUCA weights.
\end{enumerate}
In our code a WL sweep is done by a call to 
\begin{equation} \label{WLsweep}
  \tt u1\_mucabmha.f\,,
\end{equation}
which updates link variables in sequential order through calls to the 
included routine {\tt u1\_update\_mubmha.f}. This routine generalizes 
BMH updating to the situation of MUCA weights,
which is relatively straightforward (see the code for details) and 
increases efficiency compared to MUCA Metropolis simulations by a 
factor 3 to~5. It calls three modularly coded routines: The functions 
{\tt fmucaln.f} and {\tt betax.f} to calculate weights and $\beta$ 
values as needed for the BMHA. After an action update is done, the 
WL update (\ref{WLupd}) is performed by a call to the subroutine 
{\tt wala\_updt.f}, which is at the heart of our modifications of 
the WL algorithm. 

The basic point is that {\tt wala\_updt.f} does not only iterate the 
number of histogram entries, but also the mean value within each 
histogram bin. The relevant lines of that code are listed below.
\begin{verbatim}
C Put addwl <= zero in 2. part of MUCA.
        if(addwl.gt.zero) then 
          wln(ix)=wln(ix)-addwl 
          xwl(ix)=(hx(ix)*xwl(ix)+x)
        endif 
        hx(ix)=hx(ix)+one
        hx0(ix)=hx0(ix)+one
        if(addwl.gt.zero) then
          xwl(ix)=xwl(ix)/hx(ix)
        else
          xmu(ix)=xmu(ix)+x
        endif
\end{verbatim}
Besides performing the update (\ref{WLupd}) the routine tracks the 
histogram entries in the array {\tt hx} and for ${\tt addwl}>0$ the
mean value of {\tt x} within bin {\tt ix} as {\tt xwl(ix)}. For $\tt 
addwl\le 0$ MUCA simulations with fixed weights are performed. Then 
the {\tt xwl(ix)} values are kept constant, but the array {\tt xmu} 
allows one to calculate at a later stage the average within a histogram 
bin. 

The {\tt xwl(ix)} values are essential entries for the Fortran functions 
{\tt fmuca.f} and {\tt betax.f}. Logarithmic WL weights {\tt wln(ix)} 
correspond to the mean value positions. For general {\tt x} values the 
function {\tt fmucaln.f} interpolates the logarithmic weights linearly 
from the {\tt wln(ix)} weights of the two neighboring mean values:
\begin{verbatim}
      x=(E-Emin)/delE
      ix=1+int(x)
      if(x.gt.xwl(1).and.x.lt.xwl(ixmax)) then
        if(x.gt.xwl(ix)) then
          ix1=ix+1
        else
          ix1=ix-1
        endif 
        w1=abs(xwl(ix1)-x)
        w2=abs(xwl(ix)-x)
        fmucaln=(w1*wln(ix)+w2*wln(ix1))/(w1+w2)
      elseif(x.le.xwl(1)) then ...
\end{verbatim}
With this input the function {\tt betax} calculates the $\beta$
values used by the BMHA:
\begin{verbatim}
      if(x.le.xwl(ix)) then
        if(ix.eq.1) then
          betax=bmax
        else
          betax=(wln(ix-1)-wln(ix))/
     &          (xwl(ix)-xwl(ix-1))/delE
        endif
      else
        if(ix.eq.ixmax) then
          betax=bmin
        else
          betax=(wln(ix)-wln(ix+1))/
     &          (xwl(ix+1)-xwl(ix))/delE
        endif
      endif
\end{verbatim}
Although these routines are modular, to transfer the relevant arrays 
and variables into the U(1) code, they are all kept in the common block
{\tt common\_u1muca.f}.

\begin{verbatim}
C wln     MUCA logarithmic weights (w=exp(wln)).
C hx      Total count of histogram entries.
C hx0     For reconstruction of entries during 
C         one recursion segment.
C xwl     Continuously updated mean values of 
C         histogram bins 
C         (used with MUCA weights).
C xmu     Keeps track of mean values of 
C         histogram bins during 
C         fixed weights MUCA runs.
C addwl   Wang-Landau parameter.
C flat    Flatness of the histogram as measured 
C         by hist_flat.f.
C irup1   Start of WL recursion loop. 
C irec    Number of WL recursions done.
C mucarun Number of MUCA run 
C         (0 for WL recursion, then 1, 2, ...). 
C ntun    Number of tunneling (cycling) events. 
C ltun0   Logical used when incrementing ntun.
      common/wln/wln(ixmax),hx(ixmax),
     &  hx0(ixmax),xwl(ixmax),addwl,xmu(ixmax),
     &  flat,irup1,irec,mucarun,ntun,ltun0
\end{verbatim}

The common block is on the specialized U(1) level, because the array 
dimension
$$\tt ixmax = Int \left[(actmax-actmin)\,(np/(4(nd-1)))\right] $$
depends on the number {\tt np} of plaquettes of the U(1) lattice.
The relevant parameters are set by the user in {\tt u1muca.par}.

Once the system cycled from the minimum (\ref{actmin}) to the maximum
(\ref{actmax}) action value and back
\begin{equation} \label{cycling} \tt 
  actmin\ \longleftrightarrow\ actmax\ ,
\end{equation} 
a WL recursion (\ref{WLrec}) is attempted. The {\it cycling} or
{\it tunneling} condition (\ref{cycling}) ensures that the range of 
interest has indeed been covered. In addition we require that the 
sampled action histogram is sufficiently flat. The flatness is defined 
by
\begin{equation} \label{flat}
  {\tt flatness} = h_{\min}/h_{\max}\,,
\end{equation} 
where $h_{\min}$ and $h_{\max}$ are the smallest and largest numbers 
of histogram entries in the range of interest, and are calculated by 
our modular routine {\tt hist\_flat.f}. Our cut on this quantity 
is set by {\tt flatcut} in {\tt u1muca.par}. In our simulations 
\cite{BeBa06} we used ${\tt flatcut}=0.5$. This is rather weak 
compared to the requirement in the original WL approach \cite{WaLa01} 
that ``the histogram $H(E)$ for all possible $E$ is not less that 
80\%  of the average histogram $\langle H(E) \rangle$'', although their 
definition of flatness is less stringent than our Eq.~(\ref{flat}). The 
conceptual difference is that the WL paper aims at iterating all the 
way towards an accurate estimate of the inverse spectral density, 
while we are content with a working estimate, which enables 
cycling~(\ref{cycling}).  The estimate of the spectral density 
is then postponed to our continuation with frozen weight for which 
convergence of the MCMC process is rigorously proven~\cite{Bbook}, 
whereas no such proof exists for the WL algorithm.

It needs to be mentioned that the histogram {\tt hx} is accumulated 
over the entire recursion process and the same is true for the 
refinement of the weights {\tt wln}. The histogram {\tt hx0} keeps 
track of the entries between WL recursions. Presently this information 
is not used in the code. One may consider to apply a flatness criterion 
to {\tt hx0} instead of {\tt hx}. This is just one of many fine tuning 
options, which we did not explore, because the WL recursion in its 
present form took just a few percent of the CPU time in our U(1) 
simulations~\cite{BeBa06}.

The desired number of WL recursions (\ref{WLrec}) is set by the 
parameter {\tt nrec} of {\tt u1muca.par}, typically ${\tt nrec}=20$
or somewhat larger. To achieve this, {\tt nrup} (number (n) 
recursions (r) upper (up) limit) WL recursion attempts are allowed, 
each accommodating up to {\tt nupdt} WL update sweeps. The update 
sweeps are interrupted for a recursion attempt when cycling 
(\ref{cycling}) is recorded by our modularly coded subroutine 
{\tt tuna\_cnt.f}, which checks after every sweep. So, ${\tt nrup} 
\times {\tt nupdt}$ is the maximum number of sweeps spent on the WL 
recursion part. The process is aborted if the given limit is exhausted 
before {\tt nrec} WL recursions are completed.

\subsection{Fixed weights MUCA simulations and measurements}

Fixed weight MUCA simulations are performed by the routines
discussed in the previous section, only that the WL update
(\ref{WLupd}) is no longer performed, what is programmed to be
the case for ${\tt addwl}\le 0$. We still perform updates of the 
weights in-between (long) MUCA production runs, which is done by 
a call to {\tt u1mu\_rec1.f} in the initialization routine {\tt 
u1mu\_init.f}.

Overall our simulation consists of 
$$\tt nequi+nrpt\times nmeas\times nsw$$
BMHA sweeps, each (optionally) supplemented by 2 overrelaxation sweeps.
During a MUCA simulation several physical quantities (described below
in the section on data analysis) are measured by the
\texttt{u1sf\_measure} subroutine.

Measurements are performed every \texttt{nsw} sweeps and 
accumulated in the arrays defined in
\texttt{common\_u1.f}:
\begin{verbatim}
c aphase(nd,ms): Phase of the U(1) "matrix".
c                (We store aphase and 
c                the matrix on the link is 
c                e^{i aphase}.)
c act:           Energy (action) variable.
c amin,amax:     Action  act  minimum and 
c                maximum of the MC sweep.
c acpt:          Counts accepted updates.
c tsa:           Time series action array.
c a_min,a_max:   Action minimum and maximum for
c                a series of sweeps.      
c tsws,tswt:     Time series for lattice 
c                average spacelike and timelike 
c                Wilson plaquette loops.
c plreal,plimag: Space arrays for Polyakov 
c                loops in (nd-1) dimensions.
c tspr,tspi:     Time series for lattice 
c                average Polyakov loops.
c isublat:       sublattice for Polyakov loops
c                (in t=0 slice).
      common /u1/ aphase(nd,ms),act,amin,amax,
     & acpt,tsa(nmeas),a_min,a_max,tsws(nmeas),
     & tswt(nmeas),plreal(mxs),plimag(mxs),
     & tspr(nmeas),tspi(nmeas),isublat(mxs)
\end{verbatim}
and in \texttt{common\_sf.f}:
\begin{verbatim}
C Arrays for structure function measurements.
      common/u1sf/ tssf(ntotalsfbox,nmeas),
     &     nsfcomp(1:ntotalsfbox,0:ndimsf)
\end{verbatim}
Then each $\tt nmeas\times nsw$ sweeps (i.e., on each iteration of 
the \texttt{nrpt} loop) the arrays with measurements are written on 
disk by the \texttt{u1mu\_rw\_meas} subroutine, which can also read
data. With a call to {\tt u1wl\_backup.f} the current state of the 
lattice is backed up on disk. This allows one to restart the program 
from the latest iteration of the \texttt{nrpt} loop if it gets 
interrupted and ensures that not more than \texttt{1/nrpt} of the 
total running time is lost in such a case. Typically, we set 
\texttt{nrpt=32}.

\subsection{Data analysis}

The program {\tt ana\_u1mu.f} calculates action, specific heat and
two Polyakov loop susceptibilities, the program {\tt sfana\_u1mu.f} 
Polyakov loop structure factors. Definitions are given in the following 
with names of corresponding gnuplot driver files in parenthesis. The 
specific heat ({\tt plot\_C.plt}) is
\begin{equation} \label{Cbeta}
  C(\beta)\ =\ \frac{1}{N_p} \left[\langle S^2\rangle - \langle S
  \rangle^2\right]~~{\rm with}~~N_p=\frac{d\,(d-1)}{2}\,N_s^3\,N_{\tau}
\end{equation} 
where $S$ is the action ({\tt plot\_a.plt}). Besides the 
action we measure Polyakov loops and their low-momentum structure 
factors. Polyakov loops $P_{\vec{x}}$ are the $U_{ij}$ products along 
the straight lines in $N_{\tau}$ direction. For U(1) LGT each 
$P_{\vec{x}}$ is a complex number on the unit circle and $\vec{x}$ 
labels the sites of the spatial sublattice. From the sum over all 
Polyakov loops 
\begin{equation} \label{PLoop}
  P\ =\ \sum_{\vec{x}} P_{\vec{x}}
\end{equation} 
we find the susceptibility of the absolute value $|P|$ 
\begin{equation} \label{chi}
  \chi_{\max}(\beta)\ =\ \frac{1}{N_s^3} \left[ \langle|P|^2\rangle 
                      - \langle|P|\rangle^2\ \right]
\end{equation} 
({\tt plot\_CP.plt}), and the susceptibility with respect to~$\beta$
\begin{equation} \label{chi_beta}
  \chi^{\beta}_{\max}(\beta)\ =\ \frac{1}{N_s^3} \frac{d~}{d\beta}\,
  \langle|P|\rangle
\end{equation} 
({\tt plot\_CPb.plt}). The analogues in spin system are the magnetic 
susceptibilities with respect to an external magnetic field and with 
respect to the temperature.

\subsubsection{Structure factors}

Structure factors are defined by 
\begin{equation} \label{sf}
  F(\vec{k})=\frac{1}{N_s^3} \left\langle\left|\sum_{\vec{x}}
  P(\vec{x})\, \exp(i\vec{k}\vec{x}) \right|^2\right\rangle\,,
  ~~\vec{k} = \frac{2\pi}{N_s}\vec{n}\,,
\end{equation} 
where $\vec{n}$ is an integer vector to which we refer as momentum in 
the following (it differs from the structure factor momentum $\vec{k}$ 
by a constant prefactor) to describe how we average over different 
momenta.

In \texttt{sf.par}
\begin{verbatim}
C   (0,0,1), (0,1,0), (1,0,0), and so on, 
C   are stored separately.
C ndimsf: number of dimensions for 
C         the structure function.
C nsfmax: maximum value of 
C         \vec{n}=(n1,n2,n3,...,n_ndimsf) 
C         with 0 =< n1,n2,..n_ndimsf =< nsfmax.
C nsfbox: total number of \vec{n} components.
      parameter(ndimsf=nd-1,nsfmax=1,
     &  nsfbox=(nsfmax+1)**ndimsf)
\end{verbatim}
the dimension \texttt{ndimsf} of the sublattice on which SFs are 
calculated is defined and one sets the maximum value of the components 
of the momentum, \texttt{nsfmax},
\begin{eqnarray}
  \vec{n}&=&(n_1,n_2,...,n_{\texttt{ndimsf}})\,, \\
  n_i&=&0,...,\texttt{nsfmax},~~i\ =\ 1,...,\texttt{ndimsf}\,.
\end{eqnarray}
As $n_i$ counting is 0-based we measure during the simulation
\begin{equation}
  \texttt{ntotalsfbox}=(\texttt{nsfmax}+1)^{\texttt{ndimsf}}
\end{equation}
SF components. Their momenta are stored in the \texttt{nsfcomp} array.
In the example of a 4D multicanonical run given in the next section 
\texttt{ntotalsfbox}=$(1+1)^3$=8. Initialization of the arrays for 
structure factor measurements is carried out by the routine 
\texttt{sf\_box\_init}, which also outputs how momenta are initialized 
and numbered:
\begin{verbatim}
sf_box_init: Structure factors initialized:
    ndimsf =     3
    nsfmax =     1
    nsfbox =     8
Integer vectors generated:
      #    n^2      components....
      1      0      0      0      0
      2      1      0      0      1
      3      1      0      1      0
      4      2      0      1      1
      5      1      1      0      0
      6      2      1      0      1
      7      2      1      1      0
      8      3      1      1      1
sf_box_init done.
\end{verbatim}
These eight SFs are measured during this simulation. For a spatially 
symmetric lattice SF components with permuted momenta, i.e.  $(0,0,1)$, 
$(0,1,0)$ and $(1,0,0)$ are equivalent and there are only
\begin{equation}
  \texttt{nsfdiff}=\frac{(\texttt{nsfmax}+\texttt{ndimsf})!}
  {\texttt{nsfmax}!\texttt{ndimsf}!}
\end{equation}
different modes. In the example \texttt{nsfdiff}=4 and they are 
$$(0,0,0)\,,~~(0,0,1)\,,~~(0,1,1)~~{\rm and}~~(1,1,1)\,.$$
To average SFs over permutations of momenta one needs to identify 
momenta that differ up to permutations, calculate their multiplicity 
(i.e., the number of permutations) and construct a mapping from all 
momenta to the set of non-equivalent momenta. For this purpose the 
\texttt{sf\_box\_shuffle.f} subroutine is used. It returns three arrays 
corresponding to the elements described above: \texttt{nsfcomp\_diff}, 
\texttt{nsfmulti}, \texttt{nsfmapping}. Using them the analysis
program \texttt{sfana\_u1mu.f} averages SF components and outputs them 
in files prefixed with the non-equivalent momenta components (for 
instance, the SF with $\vec{n}=(0,1,1)$ is output in {\tt 
011sf006x004tmu01.d}). The SF normalization in (\ref{sf}) is defined
so that $F(\vec{k})=1$ at $\beta=0$ for all momenta and dimensions. 
The output of \texttt{sf\_box\_shuffle.f} from our example run is:
\begin{verbatim}
sf_box_shuffle:
Different SF components (integer vectors):
#  multi    n^2      components....
1      1      0      0      0      0
2      3      1      0      0      1
3      3      2      0      1      1
4      1      3      1      1      1
Mapping of the components (888 separator):
1   0  0  0   888   1   888   0  0  0
2   0  0  1   888   2   888   0  0  1
3   0  1  0   888   2   888   0  0  1
4   0  1  1   888   3   888   0  1  1
5   1  0  0   888   2   888   0  0  1
6   1  0  1   888   3   888   0  1  1
7   1  1  0   888   3   888   0  1  1
8   1  1  1   888   4   888   1  1  1
sf_box_shuffle done.
\end{verbatim}
The first part of the output shows that four different SF components
were identified and in the second part the mapping from the eight
original momenta is shown.

If only partial measurements are available, one can choose the parameter 
{\tt nset} in {\tt ana\_u1mu.f} or {\tt sfana\_u1mu.f}, which is preset
to $\tt nset=nrpt$, smaller than {\tt nrpt}.

\subsubsection{Reweighting to the canonical ensemble}

The analysis programs are reweighting to the canonical ensemble. The 
simulation is performed with $\exp(\texttt{wln}(E))$ weights, which 
need to be replaced by the Boltzmann factor $\exp(-\beta E)$. Given 
a set of $N$ multicanonical configurations the estimator for an 
observable $O$ in the canonical ensemble is
\begin{equation}\label{Orew}
\displaystyle
  \langle O\rangle(\beta)=\frac{\sum_{i=1}^NO_i\exp(-\beta E_i
  -\texttt{wln}(E_i))}{\sum_{i=1}^N\exp(-\beta E_i-\texttt{wln}(E_i))}.
\end{equation}
Eq. (\ref{Orew}) involves large terms in the numerator and denominator 
that can cause an overflow. To avoid this we use logarithmic coding
as described in \cite{Bbook}. Instead of adding two numbers one 
expresses the logarithm of the sum through their logarithms. With this
strategy one effectively evaluates the logarithm of the numerator and 
denominator, which are of the same order, and exponentiates the 
difference.

The {\tt u1\_ts\_zln.f} subroutine performs the reweighting of the time 
series to a given value of $\beta$ according to Eq. (\ref{Orew}). Since 
the reweighting procedure is non-linear, one expects a  bias, which
is for {\tt nrpt} patches proportional to $\tau_{\rm int}/{\tt (nmeas*
nsw)}$. Using jackknife error bars the bias becomes reduced by a
factor 1/({\tt nrpt}-1). This is realized by the {\tt u1\_ts\_zlnj.f}
subroutine. If one is not yet satisfied, one can go on and use the
jackknife approach to estimate the bias explicitly.

\subsection{Example runs}\label{subsec_MUCAruns}

We have prepared MUCA example runs in the subfolders 
$${\tt MUCA3D04t06xb2p0}~~{\rm and}~~{\tt MUCA4D04t06xb1p1}$$
of the folder {\tt ExampleRuns}. The values of {\tt actmin} and 
{\tt actmax} in {\tt u1muca.par} are estimates from the previously 
discussed short canonical runs. The last four characters in the 
subfolder names denote the value of {\tt beta\_table} for which 
the BMHA table is calculated.

The {\tt *.d} test files left in these subfolders were obtained from 
the analysis of MUCA data obtained by the preset runs. The MUCA data 
themselves are produced in {\tt *.D} files, which have been removed, 
because they are unformatted and readability is only guaranteed on the 
platform on which they are produced. The MUCA data production goes
through three steps of individual runs. First one has to compile and 
run the program {\tt u1wl\_bmho.f}, which uses our WL recursion to 
obtain a working estimate of the MUCA weights. Subsequently two runs
of MUCA data production are performed by the program {\tt u1mu\_bmho.f}.
After each data production step one may execute the data analysis 
programs {\tt ana\_u1mu.f} and {\tt sfana\_u1mu.f}.

\begin{figure}
\includegraphics[width=\columnwidth]{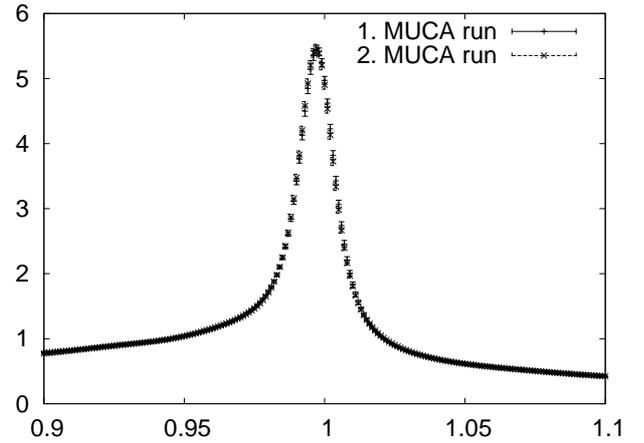}
\caption{Specific heat on a $6^34$ lattice.}
\label{fig_C}
\end{figure}
\begin{figure}
\includegraphics[width=\columnwidth]{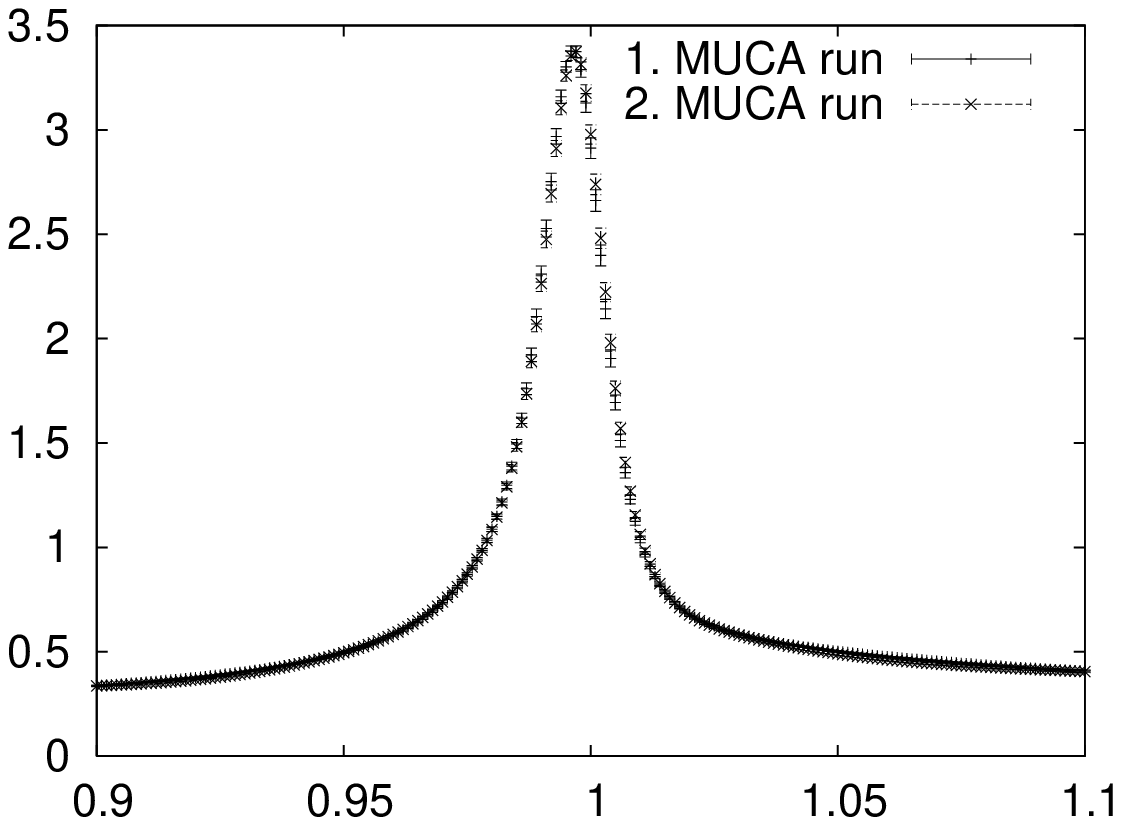}
\caption{Polyakov loop susceptibility on a $6^34$ lattice.}
\label{fig_CP}
\end{figure}
\begin{figure}
\includegraphics[width=\columnwidth]{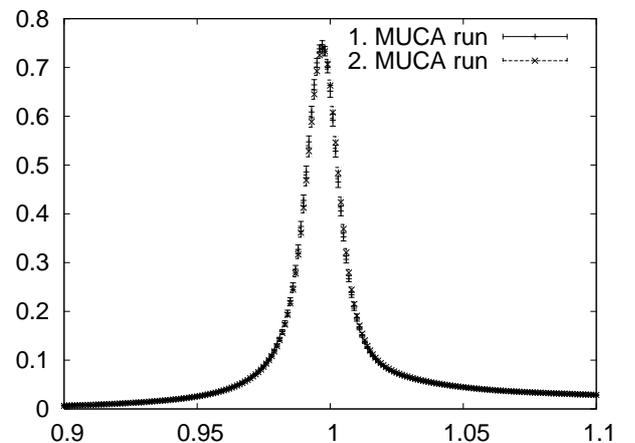}
\caption{Polyakov loop susceptibility with respect to $\beta$ on a 
$6^34$ lattice.}
\label{fig_CPb}
\end{figure}
\begin{figure}
\includegraphics[width=\columnwidth]{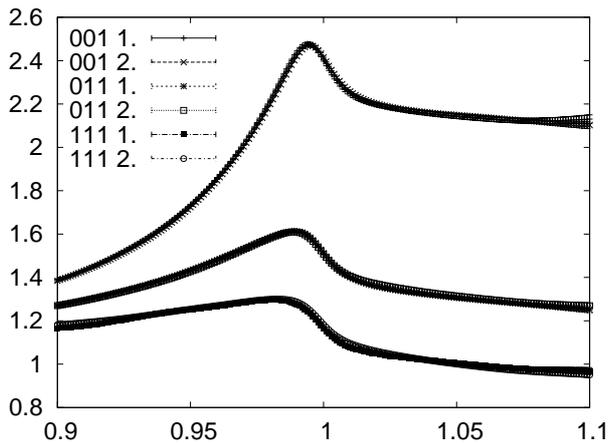}
\caption{Structure factors on a $6^34$ lattice (1.\ and 2.\ MUCA runs).}
\label{fig_sfs}
\end{figure}

In our examples the WL recursion is considered complete after 20
successful recursion steps (\ref{WLrec}). In 3D this was achieved 
after 22 cycling (tunneling) events. In 4D 23 cycling events were
needed. Then, during the simulation with fixed weights, more than
1$\,$000 tunnelings per job were recorded in 3D. In 4D 214 tunnelings 
occurred in the first and 247 in the second MUCA run. These numbers 
vary across different platforms. The results of the analysis
programs are shown in Fig.~\ref{fig_C} for the specific heat,
Fig.~\ref{fig_CP} and \ref{fig_CPb} for the susceptibilities and in
Fig.~\ref{fig_sfs} for the first three non-trivial structure factors.
On our 2~GHz PC the data production took 74m per job. Before, the 
WL recursion completed in just 2m27s.

\begin{figure}
\includegraphics[width=\columnwidth]{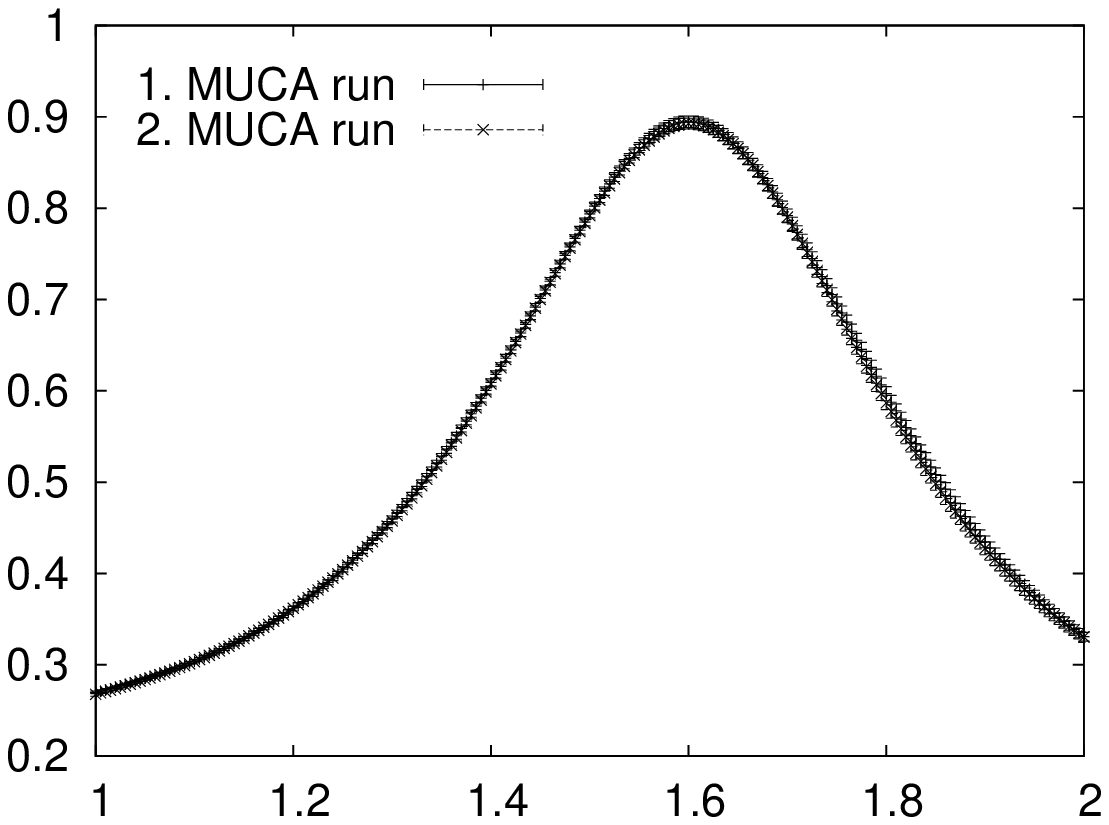}
\caption{Polyakov loop susceptibility on a $6^24$ lattice.}
\label{fig_CP3d}
\end{figure}

In 3D the specific heat does not diverge and the transition is much
broader. We show in Fig.~\ref{fig_CP3d} the Polyakov susceptibility. 
On our PC the WL recursion completed in 6s and the data production 
took 7m3s per job.

\section{Summary and Conclusions} \label{sec_conclusions}

We think that the open source Fortran code documented in this paper 
can be modified for many applications in statistical physics and LGT,
considerably beyond the U(1) gauge group.
A number of parameters can be varied, but one should have in mind 
that most of them have not been tested. 
Obviously, it is in the responsibility of the user to perform rigid 
tests and verifications before trusting any of the result. 

\section*{Acknowledgments}
This work was in part supported by the U.S. Department of Energy under
contracts DE-FG02-97ER41022, DE-FC02-06ER-41439 and NSF grant 0555397.

\bibliographystyle{elsarticle-num.bst}
\bibliography{cpc_u1}

\begin{thebibliography}{10}
\expandafter\ifx\csname url\endcsname\relax
  \def\url#1{\texttt{#1}}\fi
\expandafter\ifx\csname urlprefix\endcsname\relax\def\urlprefix{URL }\fi
\expandafter\ifx\csname href\endcsname\relax
  \def\href#1#2{#2} \def\path#1{#1}\fi

\bibitem{Rothe}
H.~Rothe, Lattice Gauge Thories: An Introduction, World Scientific, Singapore,
  2005.

\bibitem{Wi74}
K.~Wilson, Confinement of quarks, Phys. Rev. D 10 (1974) 2445--2459.

\bibitem{JeNe83}
J.~Jersak, T.~Neuhaus, P.~Zerwas, {U(1)} lattice gauge theory near the phase
  transition, Phys.\ Lett.\ B 133 (1983) 103--107.

\bibitem{ArLi01}
G.~Arnold, T.~Lippert, T.~Neuhaus, K.~Schilling, {Finite size scaling analysis
  of compact QED}, Nucl.\ Phys.\ B (Proc.\ Suppl.) 94 (2001) 651--656.

\bibitem{ArBu03}
G.~Arnold, B.~Bunk, T.~Lippert, K.~Schilling, {Compact QED under scrutiny: it's
  first order}, Nucl.\ Phys.\ B (Proc.\ Suppl.) 119 (2001) 864--866.

\bibitem{VeFo04}
M.~Vettorazzo, P.~de~Forcrand, Electromagnetic fluxes, monopoles and the order
  of 4d compact {U(1)} phase transition, Nucl.\ Phys.\ B 686 (2004) 85--118.

\bibitem{BeBa06}
B.~Berg, A.~Bazavov, Non-perturbative {U(1)} gauge theory at finite
  temperature, Phys. Rev. D 74 (2006) 094502.

\bibitem{Polyakov:1976fu}
A.~M. Polyakov, {Quark Confinement and Topology of Gauge Groups}, Nucl. Phys.
  B120 (1977) 429--458.

\bibitem{Gopfert:1981er}
M.~Gopfert, G.~Mack, {Proof of Confinement of Static Quarks in
  Three-Dimensional U(1) Lattice Gauge Theory for All Values of the Coupling
  Constant}, Commun. Math. Phys. 82 (1981) 545.

\bibitem{Berezinsky:1970fr}
V.~L. Berezinsky, {Destruction of long range order in one-dimensional and
  two-dimensional systems having a continuous symmetry group. 1. Classical
  systems}, Sov. Phys. JETP 32 (1971) 493--500.

\bibitem{Kosterlitz:1973xp}
J.~M. Kosterlitz, D.~J. Thouless, {Ordering, metastability and phase
  transitions in two- dimensional systems}, J. Phys. C6 (1973) 1181--1203.

\bibitem{Borisenko:2008sc}
O.~Borisenko, M.~Gravina, A.~Papa, {Deconfinement transition in the compact 3d
  U(1) lattice gauge theory at finite temperatures}, J. Stat. Mech. 2008 (2008)
  P08009.

\bibitem{BeNe91}
B.~Berg, T.~Neuhaus, Multicanonical algorithms for first order phase
  transitions, Phys. Lett. B 267 (1991) 249--253.

\bibitem{Bbook}
B.~Berg, {Markov} {Chain} {Monte} {Carlo} Simulations and Their Statistical
  Analysis, World Scientific, Singapore, 2004.

\bibitem{WaLa01}
F.~Wang, D.~Landau, Efficient, multiple-range random walk algorithm to
  calculate the density of states, Phys. Rev. Lett. 86 (2001) 2050--2053.

\bibitem{BaBe05}
A.~Bazavov, B.~Berg, Heat bath efficiency with a {Metropolis}-type updating,
  Phys. Rev. D 71 (2005) 114506.

\bibitem{Ad81}
S.~Adler, Over-relaxation method for the {Monte} {Carlo} evaluation of the
  partition function for multiquadratic actions, Phys. Rev. D 23 (1981)
  2901--2904.

\bibitem{Me53}
N.~Metropolis, A.~Rosenbluth, N.~Rosenbluth, A.~Teller, E.~Teller, Equation of
  state calculations by fast computing machines, J. Chem. Phys. 21 (1953)
  1087--1092.

\bibitem{We89}
R.~Wensley, Monopoles and {U(1)} lattice gauge theory, {PhD} dissertation,
  University of Illinois at Urbana-Champaign, Department of Physics,
  {ILL-TH-89-25} (1989).

\bibitem{HaNa92}
T.~Hattori, H.~Nakajima, Improvement of efficiency in generating random {U(1)}
  variables with {Boltzmann} distribution in {Monte Carlo} calculations, Nucl.
  Phys. B (Proc. Suppl.) 26 (1992) 635--637.

\bibitem{Ha70}
W.~Hastings, {Monte} {Carlo} sampling methods using {Markov} chains and their
  applications, Biometrica 57 (1970) 97--109.

\bibitem{Ad88}
S.~Adler, Overrelaxation algorithms for lattice field theories, Phys. Rev. D 37
  (1988) 458--471.

\bibitem{Cr87}
M.~Creutz, Overrelaxation and {Monte} {Carlo} simulation, Phys. Rev. D 36
  (1987) 515--519.

\bibitem{BBD08}
A.~Bazavov, B.~Berg, S.~Dubey, Phase transition properties of {3D} {Potts}
  models, Nucl. Phys. B 802 (2008) 421--434.

\bibitem{BeJa07}
B.~Berg, W.~Janke, Wang-{Landau} multibondic simulations for second-order phase
  transition, Phys. Rev. Lett. 98 (2007) 040602.

\bibitem{Ya08}
W.~Yang, et~al., Quantitative computer simulations of biomolecules: A snapshot,
  J.\ Comp.\ Chem. 129 (2008) 668--672.

\end{thebibliography}

\clearpage
\end{document}